\documentclass{emulateapj}
\newcommand{\etal}{{\it et al.}}
\newcommand{\kms}{km~s$^{-1}$}
\usepackage{lscape}

\begin{document}

\title{A Digital Archive of HI 21 cm Line Spectra of Optically-targeted Galaxies}

\author{Christopher M. Springob\altaffilmark{1}, Martha P. Haynes\altaffilmark{1,2}, Riccardo Giovanelli\altaffilmark{1,2}, and Brian R. Kent\altaffilmark{1}}

\altaffiltext{1}{Center for Radiophysics and Space Research,
Cornell University, Space Sciences Building, Ithaca, NY 14853; 
springob@astro.cornell.edu, haynes@astro.cornell.edu, riccardo@astro.cornell.edu, bkent@astro.cornell.edu}
\altaffiltext{2}{National Astronomy and Ionosphere Center, Cornell University, 
Ithaca, NY 14853.  The National Astronomy and Ionosphere Center is operated by 
Cornell University under a cooperative agreement with the National Science 
Foundation.}

\hsize 6.5 truein

\begin{abstract}
We present a homogeneous compilation of HI spectral parameters extracted from
global 21 cm line spectra for some 9000 galaxies in 
the local universe (heliocentric velocity $-200 < V_{\odot} < 28,000$ \kms) 
obtained with a variety of large single dish radio telescopes but
reanalyzed using a single set of parameter extraction
algorithms. Corrections to the observed HI line flux for source extent
and pointing offsets and to the HI line widths for instrumental
broadening and smoothing are applied according to model estimates
to produce a homogenous catalog of derived
properties with quantitative error estimates. Where the redshift is
available from optical studies, we also provide flux measurements for an 
additional 156 galaxies classified as marginal HI detections and
rms noise limits for 494 galaxies classified as
nondetections. Given the diverse nature of the observing programs
contributing to it, the characteristics of the combined dataset are
heterogeneous, and as such, the compilation is neither
integrated HI line flux nor peak flux limited. However, because of
the large statistical base and homogenous reprocessing, the spectra
and spectral parameters of galaxies in this optically targeted 
sample can be used to complement data obtained at other wavelengths
to characterize the properties of galaxies in the local universe and
to explore the large scale structures in which they reside.
\end{abstract}


\section {Introduction}

Neutral hydrogen constitutes a significant component of the
interstellar medium of disk galaxies.  
The study of its abundance and kinematics, as measured by the 21 cm emission line, has seen 
applications to a variety of astrophysical problems, ranging from studies of the large
scale distribution of galaxies to secular evolution in clusters.  HI line emission carries 
information about the distribution of neutral gas, and through tracing
of its velocity field, of the total mass in a galaxy.  Even in the absence of
any spatial information about the distribution of the gas, the global HI line
profile, contained within the beam of a single dish telescope or
integrated over the entire HI distribution where resolved, provides three key
indicators: (1) the observed redshift of the system as a whole; (2) the observed
HI line flux, which, in the optically thin case, can be related to the
HI mass; and (3) the observed HI line full width, which, when corrected to
face-on viewing, is related to the total rotational velocity of the HI
gas. These quantities have been used for diverse applications such as
the tracing of large scale structure by gas-rich systems, the comparative HI content
of galaxies in differing environments and the derivation of secondary distances
using the Tully-Fisher method (Tully \& Fisher 1977).

HI line systemic velocities, fluxes and characteristic velocity widths
have been recorded for more than 20,000 galaxies and
presented in large published compilations such as that of 
Huchtmeier \& Richter (1991) and the {\it Third Reference Catalog of Bright Galaxies} 
(RC3, de Vaucouleurs \etal ~1991). Until recently, the vast majority of
extragalactic HI line spectral parameters were associated with
observational targets selected from optical catalogs. Only recently
have large-area HI blind surveys been undertaken, notably
the Arecibo HI Strip Survey (AHISS; Zwaan \etal ~1997), the Arecibo
Dual Beam Survey (ADBS;  Rosenberg \& Schneider 2000), the 
HI Jodrell All-Sky Survey (HIJASS; Lang \etal ~2003) and the
HI Parkes All-Sky Survey (HIPASS; Barnes \etal ~2001; 
Koribalski \etal ~2004; Meyer \etal ~2004).
Most recently, the blind HI survey
conducted with the Parkes Telescope multifeed array (HIPASS)
has catalogued 4315 extragalactic HI line sources 
(HIPASS catalog; Meyer \etal ~2004) south of Dec. $< +2^\circ$ identified purely on
the peak flux of their HI line emission profiles. Perhaps surprisingly,
extremely few systems detected in the HI blind surveys have no optical
counterpart (Kilborn \etal ~2000; Rosenberg \& Schneider
2002), but understanding the difference between optical-- and
HI--selected samples is critical to determining the true census of gas-bearing
objects in the nearby universe and the relationship of present day
disks to the damped Lyman $\alpha$ absorbers seen in QSO spectra.

With the exception of the HIPASS dataset which has been processed
uniformly and can be accessed digitally through a 
public website (http://hipass.aus-vo.org/), a significant disadvantage of past
compilations arises from heterogenous processing and parameter
extraction algorithms, as well as the lack of availability of the digital
spectra themselves. In many cases, error or noise estimates are
lacking and corrections for instrumental effects not applied or
derivable. Because of our own on-going work related to the
applications of HI spectral data in correlative and secondary studies,
we have developed and maintained a digital archive of HI line spectra 
for some 9000 optically--selected galaxies observed
by us and our numerous collaborators using various telescopes over the
last 20$+$ years. Two of the telescopes used, the 91~m and 42~m
telescopes of the National Radio Astronomy Observatory{\footnote{
The National Radio Astronomy Observatory is operated by
Associated Universities, Inc. under a cooperative
agreement with the National Science Foundation.}} at Green Bank, are
no longer operational (in fact, the 91~m telescope no longer exists). 
While HI parameters for most of these datasets have been presented
previously, changing scientific objectives and
computer capabilities have led to advances in the robustness of
parameter extraction algorithms. In order to establish a well-characterized
and permanent digital archive of these global HI profiles, we have
reanalyzed the global HI line profiles using a single set of
processing and parameter estimation algorithms. In this paper, we
present a homogeneous compilation of the extracted HI line parameters
and provide quantitative corrections and error estimates.
These data are being used in a variety of 
ongoing studies, including a robust estimation of the HI mass function 
(Springob, Haynes, \& Giovanelli 2005) and studies of the local
peculiar velocity field.

In Section 2, we review the observational circumstances of the
archival dataset as well as present new HI observations undertaken
with the Arecibo 305 meter telescope.  
Section 3 discusses the process of extracting useful parameters
from the global HI line profiles and the corrections required to
derive physically meaningful quantities.  The archive is presented in
Section 4, followed by discussion of the characteristics of the data 
set, its completeness and its limitations in Section 5. A brief
summary concludes in Section 6.

\section {Observations}
The majority of the HI line datasets included in the current compilation
have been discussed in previous publications. Here we review the
general characteristics of the telescopes employed to conduct the
observations, summarized in Table 1,
and also present the results of new observations for a
sample of 145 obtained with the upgraded Arecibo telescope.

\begin{deluxetable*}{lccccl}
\tablewidth{0pt}
\tablecaption{Characteristics of Single Dish Telescopes and their Datasets\label{Tab1}}
\tablehead{
\colhead{Telescope/Feed} & \colhead{Aperture}  &
\colhead{Gain} & \colhead{HPBW} & \colhead{\# spectra} & 
\colhead{Original Reference}\\
\colhead{} & \colhead{(m)} & \colhead{(KJy$^{-1}$)} &
\colhead{(\arcmin)} & \colhead{} & \colhead{}
}                   
\startdata
Arecibo  ``dual circular''  & 305 & ~8.0~  & ~3.3 & 6475 &
HG84,a,b,c,d,e,f,g,h,k,m,n,o,p,q,s\nl
Arecibo   ``flat''           & 305 &  ~6.0~  & ~3.9 &  ~~20 & HHG,o,q\nl
Arecibo   ``Gregorian''      & 305 & 11.0~  & ~3.5 &  ~116 & this work\nl
Effelsberg                   & 100 & ~1.5~ & ~8.8 &   ~~66 & m\nl
Green Bank 91~m              & ~91  & ~1.1~  & 10.0 & 1090 & i,j,m,n,o\nl
Green Bank 42~m              & ~42  & ~0.25  & 21.0 & ~706 & l,m,r\nl
Nan\c cay               & 300$\times$34.6  & ~1.2~ & 4$\times$20 & ~377 & m\nl
\enddata
\tablenotetext{a}{Freudling \etal ~1988}
\tablenotetext{b}{Freudling \etal ~1992}
\tablenotetext{c}{Giovanelli \etal ~1986}
\tablenotetext{d}{Giovanelli \& Haynes 1989}
\tablenotetext{e}{Giovanelli \& Haynes 1993}
\tablenotetext{f}{Giovanelli \etal ~1997}
\tablenotetext{g}{Haynes \& Giovanelli 1986}
\tablenotetext{h}{Haynes \& Giovanelli 1991a}
\tablenotetext{i}{Haynes \& Giovanelli 1991b}
\tablenotetext{j}{Haynes \etal ~1988}
\tablenotetext{k}{Haynes \etal ~1997}
\tablenotetext{l}{Haynes \etal ~1998}
\tablenotetext{m}{Haynes \etal ~1999}
\tablenotetext{n}{Magri \etal ~1988}
\tablenotetext{o}{Magri 1994}
\tablenotetext{p}{Scodeggio \etal ~1995}
\tablenotetext{q}{van Zee \etal ~1995}
\tablenotetext{r}{van Zee \etal ~1997}
\tablenotetext{s}{Wegner \etal ~1993}
\end{deluxetable*}

\subsection{Archival Datasets}

Since the early 1980's, we have attempted to maintain a digital
archive of the extragalactic HI global profiles we and our
collaborators have observed at a number of large single-dish radiotelescopes,
namely the 305~m Arecibo telescope of the National Astronomy and Ionosphere Center,
the late 91~m  and 42~m Green Bank telescopes of the National Radio Astronomy
Observatory, the Nan\c cay radio telescope of the Observatory of Paris
and the Effelsberg 100~m telescope of the Max Planck Institut
f\"ur Radioastronomie. The basic parameters of those telescopes are
summarized in Table 1, along with a compilation of the principal
references to the published works
which include discussions of the individual observing programs,
observational setups and target selection criteria. Observations at
Arecibo employed several diffferent L--band feed systems, denoted by their designated
names: one of the high gain dual circular polarization feeds
(``dual circular''; two such similar feeds were available,
one of which was remotely tunable in frequency) and a low sidelobe linear
polarization L-band feed (``flat'' feed). During the 1990's, the
Arecibo telescope was upgraded and a low noise, broadband secondary and
tertiary reflector Gregorian optical system was installed; observations 
made with it are described in the following subsection.
Nearly all observations were conducted in total power
mode, with a position switching technique used to remove instrumental
baseline effects. At each telescope, calibration of the noise diodes was
performed by conducting observations of continuum sources from the catalog of
Bridle \etal ~(1972), and both the flux and velocity scales were further checked
by observations of a sample of strong HI line emitting galaxies. Typical
on--source integration times were dictated by signal--to--noise requirements
and varied from as little as 5 minutes at Arecibo to as much as 2 hours with
the 42~m and Nan\c cay telescopes. Final spectra are the accumulated weighted
averages of multiple ON-OFF pairs. 
                                                    
The main distinctions among the HI datasets
arise from differences in the sensitivities achieved with the different
telescopes, and differences in the resolution and bandwidth associated with the
available spectral backends. In order both to permit the analysis
of instrumental effects and to measure widths using the same
algorithm, the accumulated and calibrated final
spectra were exported from the respective on--line data format into the
standard ANALYZ--GALPAC ``history file'' format developed by RG at the Arecibo
Observatory and were thus available in digital form for reprocessing
using the algorithms described below. All accumulated spectra have been re-smoothed
and re-baselined using an interactive process that allows the user to select the
degree of smoothing (typically Hanning, but sometimes also a 3 channel
BOXCAR in order to improve signal-to-noise, as noted in Table 3), choose the
baseline regions and order (typically 3), flag the boundaries for flux integration
and examine the results before they are stored away in the permanent
archive.  

All spectra presented here have been reprocessed using the same
parameter extraction algorithms, producing new measurements which
supercede those presented previously.
           
\subsection{New Arecibo Observations}
In addition to the HI spectra already contained in the digital
archive, new HI line observations were conducted with the Arecibo 305
meter telescope and its Gregorian optics system between September 
and November 2001. 116 galaxies were detected, and an additional 29 galaxies for 
which we have optical heliocentric velocities yielded upper limits.  Targets were selected 
from among inclined spirals included in our private database, referred
to as the Arecibo General Catalog (AGC). The new observations preferentially
targeted those objects which appeared to be good Tully-Fisher
candidates and for which we did not already have HI spectral parameters, in the region
of the Pisces--Perseus supercluster, that is $22^h <$ R.A. $< 4^h$,
$0^{\circ} <$ Dec. $< +38^{\circ}$.

The L-band narrow
receiver was used in 9-level sampling mode for all observations, feeding each polarization to
four correlator boards with 1024 channels and 25 MHz bandwidth (yielding a velocity resolution of 5.15 \kms).  
In most (but not all) cases, the redshift was not known in advance so
that a search mode was employed, staggering the four boards to cover
a total of 85 MHz bandwidth. The first
board was centered at the 21 cm line rest frequency of
1420.40578 MHz, while the other three boards overlapped but with 
center frequencies offset at 20, 40, and 60 MHz below this frequency.  On all nights, the system temperature was approximately 30 K.  All
observations were conducted in a total 
power, position switching mode as a sequence of ON--source and
OFF-source pairs of 4 minutes each. The final spectrum is the average of the
combination of the successive ON-OFF pairs.  The 
total number of ON/OFF observations made for each galaxy varied based on the signal-to-noise 
ratio, and ranged from one to four.  On the first three nights of observations, each board was Doppler corrected to heliocentric velocity at the start of each ON-OFF pair.  Beginning with the fourth night, the same Doppler correction was applied to each of the four boards, allowing the spacing between correlators to remain at 20 MHz exactly.  Beginning on the fifth night, we used the NAIC ``radar blanker'', which detects the San Juan airport's radar and suspends data-taking for brief intervals around the pulse arrival times, for all observations in which the redshift of the galaxy was either not known a priori or known to be such that the redshifted 1420 MHz HI line would overlap with the 1350 MHz frequency of the airport radar.

Whereas data reduction for all other Arecibo observations presented here was performed 
using the Arecibo ANALYZ-GALPAC analysis system, the data reduction for these new 
observations was carried out using the AIPS++ software package
\footnote{The AIPS++ (Astronomical Information Processing System is a
product of the AIPS++ Consortium.  AIPS++ is available for use under
the Gnu Public License.  Further information may be obtained from 
http://aips2.nrao.edu.}.  The reduction algorithms match those carried out for the earlier 
observations in ANALYZ-GALPAC (and discussed in Haynes \etal ~1999 and references therein) 
exactly, with one exception.  We have used the `setfreqmed' and `settimemed' functions 
within the AIPS++ `autoflag' package to eliminate individual time-frequency pixels found 
to be corrupted by radio frequency interference (RFI).  The RFI identification algorithm 
flags pixels that deviate by more than 10 times the median deviation from a sliding median 
of width 10 pixels in both the time and frequency dimensions.  The width of each pixel in 
the time domain is 1 second.

\section {HI 21 cm Line Fluxes, Systemic Velocities and Line
  Widths}

The parameters extracted from global 21 cm line profiles obtained with
single dish telescope require correction before physical quantities
can be derived from them. Here the process for conversion from
observed to corrected parameters is described in some detail.

\subsection {Integrated Line Fluxes}

Modern radio telescopes can be continuum flux calibrated to a small
fraction of their system noise, but as discussed by van Zee \etal
~(1997), the accuracy with which HI line fluxes can be measured is
much less precise. Beyond the continuum flux calibration which can
be achieved at the better than 2\% level, the accuracy with which
line fluxes can be measured depends both on baseline stability
and on corrections that must be applied to the observed values.
The HI line fluxes $S_{obs} = \int S(V)~dV$ presented here have been
measured on the smoothed and baseline
subtracted profiles by an interactive process in which the user marks
the range of correlator channels over which the flux is measured by
integrating the emission within these boundaries. The
observed line flux $S_{obs}$ must then be corrected for (1) beam attenuation, (2)
pointing offsets, and (3) HI self absorption. We discuss each
correction separately.

Since the HI distribution in a galaxy is often partially resolved by
the telescope beam, the HI line flux detected by a given telescope
underestimates the true flux by an amount that depends on how much the
source HI distribution fills, or overfills, the beam.
Since the HI distribution is generally not known, an estimate of the
beam attenuation correction is based on the adoption of a model HI
distribution and an average telescope beam power pattern. The power
patterns of the Arecibo, 42~m, 91~m and 100~m telescopes are assumed to be
Gaussians with half power beam width (HPBW) as given in Table 1. For
the Nan\c cay telescope, the beam width is assumed to be a two-dimensional
Gaussian with HPBW of 4\arcmin ~in the R.A. direction and 20\arcmin
~in the Dec. direction. The HI distribution is modeled as the sum of
two Gaussians, one of negative amplitude to produce a central hole
(Shostak 1978;  Hewitt, Haynes \& Giovanelli 1983, hereafter
HHG). Following Section IV of HHG,
the HI surface brightness distribution 
$\sigma_{H}(r)$ is assumed to be characterized by the same axial 
ratio and to lie along the same position angle as the
optical galaxy and is scaled such that $D_{70}$, the diameter encompassing
70\% of the HI mass, equals $1.3~a$, where $a$ is the optical major
axis, in arcminutes, 
taken from either the Uppsala General Catalog (Nilson 1973, hereafter 
UGC) or the ESO-Uppsala Survey of the ESO(B) Atlas (Lauberts 1982, hereafter ESO), or 
estimated by us on the POSS-I {\footnote{The National Geographic
Society - Palomar Observatory Sky Atlas (POSS-I) was made by the
California Institute of Technology with grants from the National
Geographic Society.}}. Following HHG,
the fraction $f_1$  of the source flux that is 
detected by the telescope beam is given by

\begin{equation}
f_1 = {{{\sum_{j=1,2}} \left[ {{a_j \theta_j^2}\over{\left({1+\theta_j^2/\theta_B^2}\right)^{1/2}
\;\left({1+\theta_j^2 cos^2i\;/ \theta_B^2}\right)^{1/2}}}\right]} \over
{\sum_{j=1,2} a_j \theta_j^2}}
\end{equation}

Like HHG, we assume a double Gaussian $\sigma_H$ of amplitude $a_2 
= - 0.6 a_1$ and relative extent $\theta_2 = 0.23 \theta_1$. Note that
$\theta_B$ above is not the HPBW but rather the beam extent at which
the power pattern falls by  $e^{-1}$, assuming a Gaussian response.
The inclination $i$ of the galaxy is derived from the observed optical
axial ratio $b/a$ and an assumed intrinsic axial ratio that varies
from 0.10 for the latest spirals to 0.23 for the earliest following
Section IV.e of Haynes \& Giovanelli (1984; HG84). 

The correction factor $c_1$ that 
accounts for beam attenuation is the inverse of this fraction,
$c_1 = 1/f_1 > 1$ and approaches 1 when the HI distribution is 
significantly smaller than the telescope beam. Because the HI extent
of HI--normal spirals is significantly larger than their optical dimension,
the application of this correction is very important for most HI
observations conducted at Arecibo with its relatively small HPBW ($\sim
3.5$\arcmin). For example, face-on galaxies with optical sizes equal to 
about half the HPBW (i.e., $a \sim1.7$\arcmin) have
corrections of about $25\%$; a similar edge-on HI disk requires
a correction half as large. For an object inclined by $\sim 45^\circ$
and an optical extent equal to the HPBW, the correction
amounts to $\sim$60\%. Because the HI disk can normally be traced
well beyond the optical extent, the observed HI line flux in
face-on systems as small as 0.8\arcmin ~require corrections of 5\% or
more. The other telescopes that have been employed have substantially
wider beams, and there are few galaxies observed with them for which
this correction is significant.

Additionally, if the source HI distribution is not centered in the 
telescope beam, $S_{obs}$ will be less than it would be in the absence of such
pointing offset. Departures from centering arise both from the
intrinsic accuracy of a radio telescope to center on the input
coordinates and from errors in the target coordinates relative to the true source
position, and they are important only when such offsets subtend a
substantial fraction of the beam extent. For older observations
conducted with the Arecibo line feeds,
the rms pointing accuracy of the telescope system was $\sim ~15$\arcsec~
in both R.A. and Dec. The precise amplitude of this error is not known but has
been assumed to be random and of order 5\% (HG84).
Therefore, all HI lines fluxes derived from spectra obtained at
Arecibo prior to 1993 have been multipled by a constant factor of 
$c_2 = 1.05$ to account for this systematic bias. This correction is 
not applied to other datasets.

Similarly, older datasets are also affected by the inaccuracy of the
coordinates input to position the telesope on source at the time the
observations were conducted. In such cases, the flux is also
underestimated by an amount which depends on the ratio of the 
angular displacement of the center of the telecope beam from the 
true source location to the beam extent and is important for some of the older 
Arecibo line feed data. For spectra in our digital
archive, a measure of the amount by which the flux has been
underestimated because of coordinate inaccuracy can be calculated
because the input coordinates are recorded in the digital spectra. 
To estimate this correction factor, we have performed a set of
numerical simulations mimicking observations offset
from the Arecibo line feed beam center, by convolving the beam power
pattern with template galaxy HI distributions following the method developed in HHG.
These simulations confirm that the measured flux in the simulations is relatively 
insensitive to the position angle of the galaxy (assumed to follow the
optical image), but quite sensitive to the angular displacement of
the beam center from the center of the HI disk.  Following again HHG's result that 
$D_{70} = 1.3 a$, we relate the angular displacement $\Theta$ of the
targeted position from the true position and the optical apparent diameter $a$ to
the flux correction $c_3$, as shown in Figure 1.  For each galaxy observed 
with the Arecibo line feeds for which the pointing 
offset exceeds 0.1\arcmin, we adopt the flux correction associated
with the closest match in both optical size and pointing offset shown
in Figure 1.  The figure illustrates the point that the required flux
correction can be very large even for optically small galaxies, if the
pointing offset is large.  However, there are only 183 galaxies for which 
this correction boosts the flux by a factor of $c_3 = $1.5-2.0, and
only 62 for which the flux correction is more than a factor of $c_3 = $2.

\begin{figure}
\epsscale{1.0}
\plotone{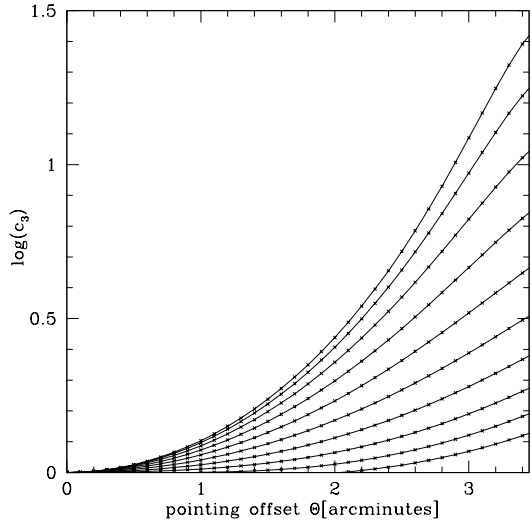}
\figcaption{Simulated pointing offset flux correction $c_3$ for older AO line feed observations for 
galaxies of different optical sizes.  The flux correction is the ratio of a galaxy's 
actual flux to its observed flux given its size and the offset of the
commanded telescope position during the observations from the optical
galaxy center.  The ten 
different curves represent corrections for different optical sizes.  The top curve is 
for galaxies with an optical major diameter of 0.54\arcmin.  The next curve for 
1.08\arcmin, and so forth, down to 5.4\arcmin. The average
pointing offset for all Arecibo observations is $\sim$ 0.21\arcmin ~so
that the application of this correction is significant only in a
relatively small number of cases. \label{FIG1}}
\end{figure}

Lastly, optical depth effects also can result in a decrease of detected
HI line flux if not all the photons escape. In the denser regions of
disks, the interstellar medium can be optically thick at 21 cm (Braun 1997), but the filling factor of opaque regions is small in
the vast majority of disks. Following Giovanelli 
\etal ~(1994) and using the optical axial ratio as the measure of 
disk inclination, we correct for HI self-absorption by multiplying
the measured HI line fluxes by the factor $c_4 = (a/b)^{0.12}$.  This 
effect is small, with less than 30\% of 
the HI line flux self absorbed in the most highly inclined systems.  However, there are some galaxies for which this is the dominant correction.  Because this correction is model-dependent, and the interested reader may wish to apply a different self-absorption correction, we report the self-absorption corrected flux separately.

The two values we report for the corrected HI line flux are then $S_c = c_1 c_2 c_3 S_{obs}$, including the corrections for source extent and pointing, and $S_c^{abs} = c_1 c_2 c_3 c_4 S_{obs}$, which also includes the correction for HI self-absorption.  $S_c^{abs}$ is, on
average, 25\% larger than the observed line flux. It should be noted
that all of the corrections increase the value of line flux used in
the calculation of HI masses. Comparison of values presented here with
others in the literature should take into account similar treatments
in order to avoid systematic differences in the derived HI masses
which might otherwise be interpreted as physical differences.

\subsubsection {Integrated Line Flux Errors}

Several different sources of error affect the measurement of
integrated HI line fluxes. The first is the uncertainty in
the absolute flux calibration commonly achieved by firing
a noise diode at regular intervals (van Zee \etal ~1997)
and understanding well the telescope gain and its variation.
Because modern telescope systems are significantly more
stable than older ones, we follow van Zee \etal ~(1997)
in estimating that the uncertainty introduced by the noise 
diode calibration is 2\% of the observed integrated flux 
for those observations made with the 42 m Green Bank 
telescope and for those made with the Arecibo telescope 
after the installation of its Gregorian optical system.  
For all other observations, we estimate the uncertainty 
introduced by the noise diode calibration to be 10\% of 
the observed integrated flux.  Further, we estimate that the uncertainty introduced by the correction coefficients is $(c_1 c_2 c_3 c_4 - 1)S_{obs}/3$.

The integrated flux error contribution from {\it statistical} uncertainty has been discussed by both Fouq\'ue \etal ~(1990b) and Schneider \etal ~(1990), both of whom derive analytical expressions for the statistical uncertainty.  Unlike the Fouq\'ue \etal ~(1990b) formula, however, the Schneider \etal ~(1990) formula does not include any assumptions regarding the shape of the line profile, but {\it does} include the contribution to the total error stemming from uncertainties in the baseline fit.  We therefore adopt the Schneider \etal ~(1990) formula, with one modification.  The formula includes a dependence on $W_{P20}$, the line width measured at 20\% of the peak flux density.  We have our own prefered width measurement algorithm which is explained in Section 3.2.  Because we find that the value of $W_{P20}/W_{F50}$ (where $W_{F50}$ is the width as measured by our favored algorithm) is, on average, $\sim 1.2$, we replace the $1.2 W_{P20}$ term in Schneider \etal ~(1990) with $1.4 W_{F50}$.  $\epsilon_{S}^{stat}$, the statistical component of the integrated line flux error, is then given by

\begin{equation}
\epsilon_{S}^{stat} = 2 ~rms \sqrt{1.4 W_{F50} \Delta V}
\end{equation}

\noindent (where $rms$ is the r.m.s. figure, measured in the signal--free part of the spectral baseline and $\Delta V$ is the velocity resolution of the spectrum).  Our estimate of the total integrated line flux error, $\epsilon_S$, is then the sum in quadrature of the error contributions from statistical noise and baseline uncertainties, noise diode calibration, and correction coefficients.  For the vast majority of the galaxies in the archive, this uncertainty is 10-15\% of the corrected integrated flux $S_c^{abs}$.

\subsection {Systemic Velocities and Velocity Widths}

The systemic velocity of the HI line profile is usually derived as the
midpoint of the segment that connects two points on opposite edges of the
emission profile at a selected intensity level, usually some fraction
of the mean or peak flux.  The same algorithm is also
used to derive the HI line width, and its extraction often dictates
the choice of measurement algorithm.
Over the years, numerous algorithms have been applied to derive
width measures that both are robust in the presence of noise and most closely reflect
the disk rotational velocity (Lewis 1983; Schneider \etal ~1986;
Chengalur \etal ~1993).
Algorithms applied originally to HI spectra contained in 
earlier datasets were optimized to produce consistent measurements in
noisy spectra (e.g., Bicay \& Giovanelli 1986) 
whereas more recent studies added emphasis on sampling the rotational
velocity (Lavezzi \& Dickey 1997).

The algorithm adopted to obtain HI line velocities and line
widths here is very similar to that discussed by Chengalur \etal
~(1993) and has been previously discussed in Haynes \etal ~(1999). 
Figure 2 illustrates the issues involved in systemic velocity and width extraction from
global HI line profiles through simulations of representative cases. 
After identification of the flux level, $f_p$, of 
each of the spectral horns, the rising side of the profile is fit by a polynomial between 
the levels of 15\% and 85\% of $f_p - rms$.  The velocity
flagged for that side of the profile $V_l$
is that for which the polynomial fit has a flux of 50\% of the value of $f_p - rms$.  This 
process is repeated for the other side of the profile to flag a
velocity $V_u$ at the corresponding level on the other horn. The
adopted velocity width, hereafter designated  $W_{F50}$,
is defined as the difference between these
two velocities  $W_{F50} = V_u - V_l$,
while the system velocity is their average $V_{21} = (V_u + V_l)/2$. In
the vast majority of cases, the polynomial fits are straight lines.  In rare cases, however, 
higher order fits are required and can be selected during the analysis stage.

\begin{figure}
\epsscale{1.0}
\plotone{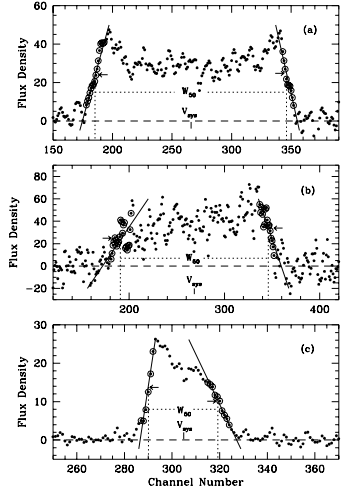}
\figcaption[] {Examples of 21 cm line profiles, illustrating the
systemic velocity and velocity width
measuring algorithm. The spectral resolution and unit, in abscissa
(``channel number'') is typically 2--11 \kms.
In panel (a), an average quality profile is illustrated,
with horns near channel numbers 196 and 340 respectively. Straight lines
are fit to the outer slopes of the profiles (circled points were used in
the fit), and the channels corresponding to the flux level equal to 50\%
of the peak level, on the fit, are flagged. Velocities $V_l$ and $V_u$
defined in the text correspond, in panel (a), to the center of channels
185 and 347. Velocity width $W_{50}$ (in spectrograph channel units) and
systemic velocity are indicated. In panel (b), a low signal--to--noise
profile is shown, illustrating some of the difficulties that can be
encountered by the algorithm in identifying the horns and fitting the
slopes. In panel (c), an asymmetric profile is shown, for which the
identification of one of the horns is also arduous, due to the structural
asymmetry of the line profile. \label{FIG2}}
\end{figure}

This procedure 
has a major advantage over width algorithms which measure a width as
the difference in velocities between points {\it on the spectrum itself} at N\% 
of the peak flux.  Under the scheme adopted here, noise effects on either side of 
the profile are averaged out.  Additionally, the use of the parameter $f_p - rms$ rather than 
simply $f_p$ substantially reduces the dependence of the measured width on the signal-to-noise ratio (SNR).
The most important flaws in the method are encountered when it is difficult
to identify the location of the horns, such as in low signal--to--noise
profiles as shown in panel 2(b), or when one of the horns is missing as
in the strongly asymmetric profile shown in panel 2(c) (see, e.g. Richter
\& Sancisi 1994 and Haynes \etal ~1998). In these cases,
the width measurement loses objectivity and accuracy and its derived
error may not reflect the true uncertainty. For this reason, we also
assign a width ``quality index'' to each result listed in Table 3.

To convert $W_{F50}$ into a physical rotational velocity, four corrections must be applied:  1) an instrumental correction that accounts for the spectrometer resolution, smoothing 
applied to the spectrum, and SNR, 2) a cosmological redshift correction,  3) a correction for turbulent motions of the HI gas, and 4) a correction for the inclination 
of the disk to the line of sight. The widths presented here account for the first two effects 
only.  The redshift correction is simply the instrumental-corrected width divided by a factor 
of $1+z$, where $z$ is the galaxy's redshift.  The instrumental correction is described in 
the following subsection. Although we do not apply such a correction,
the effect of turbulent motions on the measurement of $W_{F50}$ is
discussed in Section 3.2.3.

In addition to the width measurement algorithm described above, each spectrum has 
been measured using four additional algorithms: $W_{M50}$ is the width measured at 
50\% of the mean flux density, $W_{P50}$ is the width measured at 50\% of the value 
of the peak flux density minus the rms value $f_p - rms$, $W_{P20}$ is the width 
measured at 20\% of the value of the peak flux density minus the rms value $f_p - rms$, 
and $W_{2P50}$ is the width measured at 50\% of each of the two peaks minus the rms 
value.  We include these values to allow the interested reader to compare our width 
measurement results to those of other authors.  However, we have derived corrections 
only for $W_{F50}$, not for any of the additional width measurements.

\subsubsection{Smoothing}

It is common practice to convolve radio spectra obtained with autocorrelation spectrometers 
with a one--channel--wide cosine function (in practice a [0.25,0.5,0.25] three--channel 
function: Hanning). This is done for two reasons. First, narrowband terrestrial
interference, which occurs often within the observed spectral range
even within parts of the spectrum protected by regulatory agreements
for radio astronomical use, introduces ``ringing'' associated 
with the $(\sin x)/x$ functional response of the spectrometer to single--channel--wide 
features. And, second, because of the finite number of correlator
channels, the true autocorrelation function is truncated, resulting in
its apparent multiplication by a rectangular
function which is a multiple of the autocorrelation lag time; this in
turn means that the true power spectrum has been convolved
with the Fourier transform of this rectangular function, itself a 
``sinc function''. The application of Hanning smoothing reduces the spectral resolution 
of the observed spectrum. In some cases, further spectral smoothing is required by 
the low SNR of the observation or for other reasons; these measurements 
tend to be less useful for TF applications. All the HI spectra presented here are Hanning 
smoothed once (smoothing code H in our data tables, e.g. Haynes \etal ~1997); in a minority of cases, as noted in our tabulations, three--channel ``boxcar'' averaging plus Hanning is applied 
instead (smoothing code B).

\subsubsection{Instrumental and Noise Effects}

Previous width measurement algorithms have often relied on measuring the width by computing 
the velocity difference between channels on either side of the profile.  The channels are 
identified by either starting from inside the profile and searching outwards or starting 
outside and searching inwards, flagging the first channel to cross a flux limit that represents 
a particular fraction of the peak flux.  This procedure creates a systematic bias in the 
width measurement as a function of the SNR (Lewis 1983).  As described 
earlier, our width measurement procedure minimizes this effect by instead calculating the 
velocity difference between {\it polynomials that have been fit to several channels}.  
Nevertheless, there remains {\it some} dependence on the SNR, defined here
as the ratio of the peak HI signal to the rms noise in the line--free
portion of the spectrum, which we discuss in this section.

We have studied the instrumental and noise effect via simulations that incorporate 
autocorrelator spectrometer channel response and a variety of observed characteristics, 
such as profile slope and curvature.  The simulations involve inputting a high resolution, 
noiseless profile, then binning the flux into channels, adding noise, and measuring the 
width of the profile using the same algorithm used also on the real
data.  This process is applied to a range of simulations adopting
channel resolutions ranging from 5 to 11 \kms, for the two
different smoothing prescriptions, and for a variety of different profile shapes.

To account for the spectral noise, the true SNR of the spectrum can be characterized by 
SNR $= (f_p - {\it rms})/{\it rms}$.  The broadening of the width 
measurement is denoted by the parameter $\lambda$, which is one half the difference between the 
measured width and the width that would be measured in the absence of instrumental and 
noise effects (measured in channels).  We can then account for the combined instrumental 
and noise effects by {\it subtracting an offset} from the measured width, parametrized as 

\begin{equation}
\Delta s = 2 \Delta v ~\lambda,  
\end{equation}

\noindent where $\Delta v$ is the spectrometer channel separation in \kms~ assumed
to be equal to or very close to the channel width) and $\lambda$ 
depends on the SNR and the type of smoothing.

The simulations suggest that $\lambda$ is influenced by two competing effects.  As our 
width measurement algorithm measures a best-fit line to the sides of the profile at 50\% 
of the value of $f_p - {\it rms}$, the two parameters in play are 1) the value of 
$f_p - {\it rms}$, and 2) the range of spectral values over which the line is fit.

If instrumental and noise effects increase (decrease) $f_p - {\it rms}$, then $\lambda$ 
will decrease (increase), as a higher (lower) 50\% value corresponds to a narrower 
(broader) width.  Even in the absence of noise, the peak flux is depressed by channel 
binning and smoothing, broadening the measured width.  For realistic values of SNR, 
however, the situation is more complicated.  One can characterize $f_p - {\it rms}$ 
with the expression

\begin{equation}
f_p - {\it rms} = Max(y_i + \delta_i {\it rms}) - {\it rms},  
\end{equation}

\noindent where $y_i$ is the flux of channel $i$ that one would measure in the absence of noise 
and $\delta_i$ is a gaussian random variable describing the noise contribution to channel $i$.  
We find that profiles with steep inner slopes observed with broad channel spacings tend to 
produce decreasing values of $f_p - {\it rms}$ with increasing ${\it rms}$.  Other profile 
shapes and spectral resolutions give a somewhat more complicated dependence.  So the variation 
of $f_p - {\it rms}$ affects the profile width measurement, but whether it becomes broader 
or narrower depends on the details of the profile shape and resolution.

$\lambda$ is also influenced by the range over which the line is fit to the sides of the profile.  
The endpoints of the fit are found by searching outward from each peak, identifying the 
first channels with flux less than 15\% and 85\% of $f_p - {\it rms}$.  This method introduces 
a slight bias towards the inclusion of channels whose flux has been {\it depressed} by noise.  
For noisier profiles, the region over which the line is fit also moves closer to the peak of 
the profile, where the slope is shallower.  Both of these effects preferentially select 
channels with flux that falls {\it below} the best fit line in the noiseless case, leading 
to narrower width measurements.  This effect dominates the dependence of $\lambda$ on SNR
over the range $0.6 <$ log(SNR) $< 1.1$ for each of our channel resolutions and for all realistic 
profile shapes, meaning that $\lambda$ increases with SNR over that range.

The simulations show that the most relevant line profile parameters in determining $\lambda$ 
are the slopes on each side of each of the two peaks and the number of channels leading up to 
the peaks.  While $\lambda$ typically varies by $\sim$ 0.5 channels for the same profile shape 
with different values of the SNR, the variation can be even greater ($\sim$ 1 channel) 
among different profile shapes {\it and} among different noise realizations of the same profile 
for constant SNR, channel width, and smoothing prescription.  This variation limits our 
ability to determine $\lambda$ precisely for a real profile.  However, we can simply apply 
our simulation results for a profile with mid-range values for the inner and outer slopes 
and number of channels leading up to the peaks.  The resulting approximate solution gives:

\begin{equation}
\lambda = \lambda_{1}(\Delta v) ~~~{\rm for~~log(SNR)} < 0.6 
\end{equation}
\begin{equation}
\lambda = \lambda_{2}(\Delta v) + \lambda_{2}'(\Delta v) {\rm log(SNR) ~~~for}~~0.6 < {\rm log(SNR)} < 1.1
\end{equation}
\begin{equation}
\lambda = \lambda_{3}(\Delta v) ~~~{\rm for~~log(SNR)} > 1.1 
\end{equation}

The values of $\lambda_{1}(\Delta v)$, $\lambda_{2}(\Delta v)$, $\lambda_{2}'(\Delta v)$, and 
$\lambda_{3}(\Delta v)$ are constants for both $\Delta v < 5$ \kms ~and $\Delta v > 11$ \kms, but 
depend linearly on $\Delta v$ within the range $5 < \Delta v < 11$ \kms. The values of these 
parameters for each of the channel widths and smoothing prescriptions can be found in Table 2.  
We stress that both the linear dependence of $\lambda$ on log(SNR) in the range $0.6 <$ log(SNR) 
$< 1.1$ and the linear dependence of the $\lambda_i$'s on $(\Delta v)$ in the range $5 < \Delta v 
< 11$ \kms ~are approximate and completely empirical.  We also keep the $\lambda_i$'s constant 
outside the range $5 < \Delta v < 11$ \kms ~solely because few of our observations have channel 
resolutions far outside of that range, and we have done little testing of the width corrections 
outside that range.

\begin{deluxetable*}{rccccc}
\tablewidth{0pt}
\tablecaption{Velocity Width Instrumental Correction Parameters\label{Tab2}}
\tablehead{
\colhead{$\Delta v$}   & \colhead{$smoothing$}     & \colhead{$\lambda_{1}(\Delta v)$}   
& \colhead{$\lambda_{2}(\Delta v)$}   & \colhead{$\lambda_{2}'(\Delta v)$}    & \colhead{$\lambda_{3}(\Delta v)$}                     
}
\startdata
$<$ 5 \kms  & H  &  0.005 & -0.4685 &  0.785 &  0.395\nl
5-11 \kms  & H  &   0.037$\Delta v$ - 0.18 & 0.0527$\Delta v$ - 0.732  &  -0.027$\Delta v$ + 0.92 &  0.023$\Delta v$ + 0.28\nl
$>$11 \kms  & H  &   0.227 & -0.1523 &  0.623 &  0.533\nl

$<$ 5 \kms  & B  &   0.020 & -0.4705 &  0.820 &  0.430\nl
5-11 \kms  & B  &   0.052$\Delta v$ - 0.24 & 0.0397$\Delta v$ - 0.669 &  0.020$\Delta v$ + 0.72 &  0.062$\Delta v$ - 0.12\nl
$>$11 \kms & B  &   0.332 & -0.2323 &  0.940 &  0.802\nl

\enddata
\end{deluxetable*}

In addition to the instrumental and smoothing corrections, the observed width must
also be corrected for redshift stretch, giving:

\begin{equation}
W_{c} = (W_{F50} - \Delta_s)/(1+z) = (W_{F50} - 2 \Delta v ~\lambda) /(1+z)
\end{equation}

\subsubsection {Turbulent Motion Correction} 
 
The HI clouds in a disk do not move along strictly circular orbits. Even in the
absence of large--scale anomalies, such as warps, bars and other global
asymmetries, deviations from circular motion occur, such as those associated
with small--scale dispersion and streaming motions near spiral arms.
Bottinelli \etal ~(1983) have suggested that the contribution to the observed
width of turbulent motions, or otherwise small--scale deviations from
circular rotation, could be expressed via a term $W_t$,
additive to the circular width in the form: $W = 2 V_{max}\sin i + W_t$
where $i$ is the disk inclination and $V_{max}$ the maximum rotational
width. Tully \& Fouqu\'e (1985) found that, while a linear correction
to the observed width may well account for the effect of turbulent motions
on large widths, for low luminosity galaxies a quadratic combination would
be more appropriate (see also Fouqu\'e \etal ~1990a).  They thus proposed a formula, 
their Equation 12, that yields a linear combination at large widths, a quadratic 
combination at small widths, and a transitional form at intermediate widths.  This formula 
was adopted by a variety of other authors (e.g., Broeils 1992; Rhee 1996; Verheijen 1997) 
who also measure HI widths at N\% of the profile peak.  
These authors measure the value of the $W_t$ parameter by minimizing the scatter between the 
turbulence-corrected velocity widths and the velocity widths determined from rotation curves 
of the same galaxies.  Verheijen (1997) reviews the results of the aforementioned authors, 
and notes that the optimal value of $W_t$ strongly depends on the instrumental correction used.

Since we do not measure velocity widths at N\% of the profile peak, and our instrumental 
correction differs from that of these other authors, we are not justified in adopting 
any of the previous values of $W_t$.  Instead, we derive our own turbulent motion 
correction by introducing a random, isotropic $\sigma = 10$ \kms ~turbulent motion into 
the simulations described in the previous subsection.  We find that this depresses the 
peak flux, and thus broadens the measured profile width by moving the $50 \%$ value 
farther from the profile center.  As in Verheijen (1997) and references therein, the 
amount of turbulence-induced width broadening is found to vary greatly from galaxy 
to galaxy, depending on the details of the profile shape.  We do not find a discernable 
trend in the amount of width broadening with profile width or signal-to-noise ratio.  
For the sake of simplicity, we thus subtract the turbulence correction from the 
width linearly, with the equation: $W_{c,t}=W_{c}+W_{t}$, where $W_t =
6.5$ \kms. Lavezzi \& Dickey (1997) likewise discuss the merits of a
simple linear subtraction of a turbulence term.

As stated earlier, in order to use $W_{c}$ for applications such as
those involving the Tully-Fisher relation, corrections for both
inclination angle and turbulence must be applied as well.  Because of its uncertainty, the values of corrected width $W_c$
reported here {\it do not include that correction}, nor is viewing
angle considered. Authors wishing to make full use of the data should
apply those corrections themselves. 

\section {Data Compilation}

All of the spectra in the archive have been newly reprocessed using the new flux and width 
corrections.  The spectra themselves will be made available through the U.S. National Virtual 
Observatory.  Here, we provide the corrected and uncorrected spectral parameters.  In Table 
3, we present the results for all 8852 spectra for which we are confident that we have 
detections.  The format, described below, closely resembles that of Haynes \etal ~(1999) Table 1:

 Column (1). Entry number in the UGC, where applicable, or else in our private database, referred 
to as the Arecibo General Catalog (AGC).

     Column (2). NGC or IC designation, or other name, typically from
 the {\it Catalog of Galaxies and Clusters of Galaxies} (Zwicky \etal
 ~1961-1968), ESO, or the 
{\it  Morphological Catalog of Galaxies}
(Vorontsov-Velyaminov \& Arhipova 1968). Where used, the designation
 in the latter is abbreviated to eight characters.

     Columns (3) and (4). Right Ascension and Declination in the J2000.0 epoch either from the 
NASA/IPAC Extragalactic Database (NED) {\footnote {The NASA/IPAC Extragalactic Database is 
operated by the Jet Propulsion Laboratory, California Institute of Technology, under contract 
with the National Aeronautics and Space Administration.}}
or measured by us on the POSS-I. Typically, the listed positions have $\le 5$\arcsec accuracy.

     Column (5). The blue major and minor diameters, $a \times b$, in arcminutes, either from the 
UGC or ESO catalogs or as estimated by us on the POSS-I.

     Column (6). The morphological type code following the RC3 system. 
Classification comes either from the UGC or ESO catalogs or from our own visual examination of 
the POSS-I prints.

     Column (7). The observed integrated 21 cm HI line flux $S_{obs} = \int S~dV$ in Jy \kms.

     Column (8). The corrected integrated 21 cm HI line flux $S_c$, also in Jy \kms, after 
corrections applied for pointing offsets (Arecibo line/flat feed
spectra only) and source extent following Section 3.1. 

     Column (9). The self-absorption corrected integrated 21 cm HI line flux $S_c^{abs}$, also in Jy \kms, after 
corrections applied for pointing offsets (Arecibo line/flat feed
spectra only), source extent, {\it and} HI self-absorption, following Section 3.1. 

     Column (10). The uncertainty $\epsilon_S$ in the self-absorption corrected integrated flux $S_c^{abs}$, following Section 3.1.1.

     Column (11). The rms noise per channel of the spectrum, rms, in mJy.

     Column (12). The emission profile signal-to-noise ratio, SNR, taken as the ratio of the 
peak HI line flux to the rms noise within the signal--free portion of
the spectrum.  Unlike in Section 3.2.2, we do not subtract the rms from the peak flux before dividing by the rms.

     Column (13). The heliocentric velocity $V_{\odot}$, in \kms, of the HI line signal, taken 
as the midpoint of the profile at the 50\% level also used to measure the width (see below).

     Column (14-18). The full velocity width of the HI line in \kms, uncorrected for redshift 
or other effects, using the five measurement algorithms discussed in Section 2.2.  Widths 
are $W_{F50}, W_{M50}, W_{P50}, W_{P20},$ and $W_{2P50}$ respectively.

     Column (19). A corrected velocity width $W_c$, in \kms, which accounts for redshift 
stretch, instrumental effects, and smoothing. The correction is applied to $W_{F50}$.  Note that 
this width is not rectified for viewing angle or turbulence.

     Column (20). The estimated error on $W_c$, $\epsilon_w$, in \kms, taken to be the sum 
in quadrature of the measurement error and the estimated error in broadening corrections. 
The estimated error on the velocity itself, $\epsilon_{V}$, can be approximated as $\epsilon_{w} / 2^{1/2}$.

     Column (21). A code indicating the instrument used: (AOlf) Arecibo, line feed system; 
(AOff) Arecibo, flat feed system; (AOG) Arecibo Gregorian; (Effs) Effelsberg 100 m; (GB300) 
Green Bank 91 m; (GB140) Green Bank 42 m; (Nanc) Nancay.

     Column (22). Bandwidth of the spectrum in kHz $\Delta f$.

     Column (23). Number of channels in the spectrum $N_{chan}$.

     Column (24). Three codes indicating the smoothing, width quality, and inclusion in the 
2001 Arecibo observations.

     The first code is the smoothing code: (H) single Hanning only; (B) Hanning plus three-channel boxcar.

     The second is a qualitative assessment of the quality of the profile: (G) good; (F) fair; 
(S) single peak; (P) poor quality for TF applications; (C) confused; (A) shows absorption. Velocity 
widths measured from profiles classified as good detections should be useful for TF applications; 
those measured on profiles denoted ``F'' should be used with caution. Because the width measuring 
algorithm is designed for application to two-horned profiles, the widths measured on single peaked 
profiles should also be used with caution, as they may underestimate the true rotation width. Widths 
in the last two categories cannot be used for TF purposes.  Confusion is identified only when 
contamination from another galaxy in the beam is certain.

The third code where applicable indicates, with a '*', inclusion in the sample observed with the 
Arecibo Telescope in the fall of 2001 as described in Section 2.2.

In addition to the HI detections presented in Table 3, Tables 4 and 5
also summarize useful data about objects for which homogeneous upper limits to the
HI flux can be calculated. Table 4 contains summary data for 156
galaxies classified as ``marginal'' detections, that is, for those cases in which the signal 
has not been verified through adequate reobservation.  Table columns match those of Table 3, except 
that we omit Columns 14-20 because of their uncertainty, and Column 24
because, in this instance, all spectra have been both Hanning and
Boxcar smoothed. Table 4 also adds Column 25, the heliocentric 
velocity as derived from optical observations, $V_{\odot - opt}$, in \kms, taken from 
the NED. Although the parameters extracted from the HI line profile
are uncertain, the coincidence of the HI signal with the expected
location in frequency corresponding to the optical redshift brings a
measure of believability to the possible HI signal. Such objects may
be deserving of further, deeper HI observations which probe
sensitivities significantly greater than those presented here.

In Table 5, we present summary results for objects which are classified
as the HI line nondetections, except that Columns 7, 8, 9, 10, 12, and 13 are
also omitted. Column 26 adds the heliocentric velocity $V_{cent}$ corresponding
to the central frequency of the band searched for HI
emission. Only objects for which the radio frequency band covered by
the spectrum included the frequency expected from the optical redshift
are listed in this table. Upper limits to the HI line flux can be
estimated from the rms noise per channel and an assumption of the HI
profile shape. A simple estimate of the upper limit
to the integrated HI line flux, useful for calculating an upper
limit to the HI mass for non-detections, adopts a rectangular profile 
of height 1.5~rms~$\delta W$, where $\delta W$ is the full width of the Doppler
broadened HI line, expected, for example, for a galaxy of a given optical
luminosity given by the Tully--Fisher relation.

\section {Dataset Characteristics and Limitations}

In contrast to a wide area, HI blind survey like HIPASS, the archival
dataset presented here is extremely heterogenous in terms of its
sky coverage, depth and completeness. At the same time, it contains
HI line spectra for more than twice as many detections as the HIPASS
HICAT (Meyer \etal ~2004), with much higher typical sensitivity and
probing a greater volume.

The sky distribution for all detected galaxies listed in Table 3 is shown in Figure 3. 
The unevenness of sky coverage, with strong preference for that
portion of the sky visible to the Arecibo telescope, is clearly
evident. In fact,  77 \% of the objects fall within the Arecibo declination limits 
($-2^\circ <$ Dec. $< +38^\circ$). The southern cutoff at Dec. $\sim
-40^\circ$ represents the limitations of the steerable northern hemisphere
telescopes that were used.
There is also a strong deficiency of objects in the Zone of
Avoidance, not just because the archive is comprised of objects with
known optical counterparts but also because several of the previous surveys
have avoided regions of high galactic extinction for reasons 
of photometric accuracy and completeness at optical wavelengths.

\begin{figure}
\epsscale{1.0}
\plotone{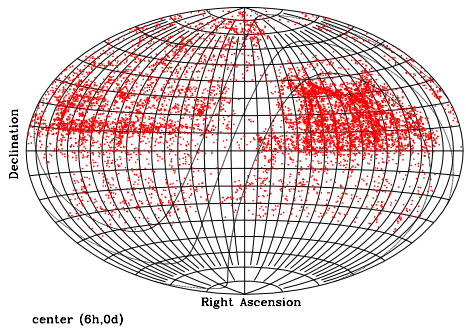}
\figcaption{Aitoff equal area projection of the
sky distribution of galaxies in the HI archive.  Shown here
are all galaxies with reliable HI fluxes and widths (all those listed 
in Table 3).  The plot is centered at R.A. = $6^h$.  The thick lines 
trace the galactic latitudes $b=-20^\circ$, $b=0^\circ$, 
and $b=+20^\circ$.\label{FIG3}}
\end{figure}

The sky distribution of targeted galaxies also reflects the large
scale structure within the survey volume, evident both in Figure 3 and
in the redshift distribution illustrated in Figure 4. The 
most prominent feature is the Pisces-Perseus Supercluster at $22^h <$
R.A. $< 4^h$ and $cz \sim 5000$ \kms which has been the target of
several past observing programs. Also seen clearly is the Virgo
region toward the center of the Local Supercluster at R.A. $\sim
12^h$,  $cz \sim1500$ \kms.

\begin{figure}
\epsscale{1.0}
\plotone{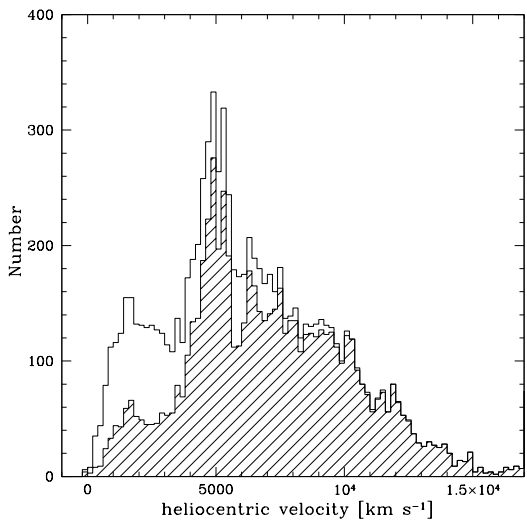}
\figcaption{Distribution of heliocentric radial velocities for all 
galaxies in the archive, in bins of width 200 \kms.  Observations
conducted at Arecibo are indicated by the shaded histogram.  The impact of
local large scale structures and the preferential coverage
corresponding to that part of the celestial sphere visible to the
Arecibo telescope, is evident. \label{FIG4}}
\end{figure}

The distributions of the rms noise per channel
and SNR are shown in Figures 5 and 6 respectively. Because of the 
superior gain and angular resolution of the
Arecibo telescope, observations undertaken with it are typically more
sensitive than ones conducted with other instruments. For this reason,
the fraction of the archival dataset extracted from Arecibo
observations is indicated separately in those figures.
The diversity of the observational
programs which have contributed datasets to this HI archive is evident
in these distributions. In some instances, the targetting of objects
for the explicit purpose of measuring comparative HI content or
accurate HI line widths has placed more stringent demands for lower
rms noise values, or higher SNR, than, for example, those programs
that were conducted principally as wide area redshift surveys.

\begin{figure}
\figurenum{5}
\epsscale{1.0}
\plotone{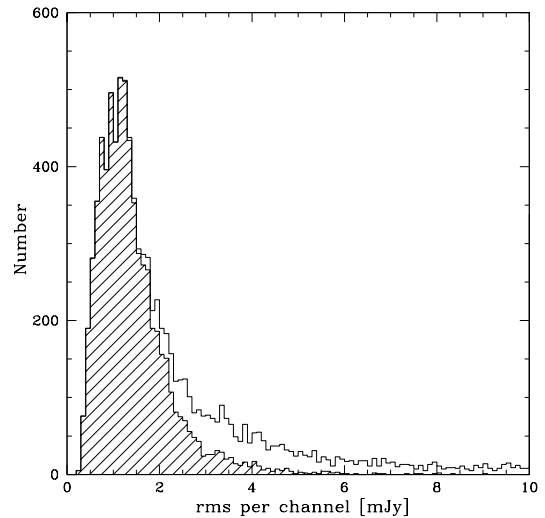}

\figcaption{Distribution of rms noise values per channel for all 
galaxies in the archive, in bins of width 0.1 mJy. Observations
conducted at Arecibo are indicated by the shaded histogram. \label{FIG5}}
\end{figure}

\begin{figure}
\figurenum{6}
\epsscale{1.0}
\plotone{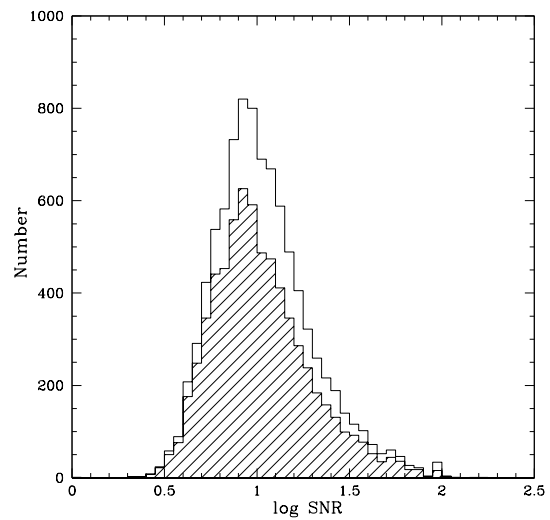}

\figcaption{Distribution of peak--signal--to--rms noise ratios, SNR, for all galaxies in 
the archive, in bins of width 0.05 dex. Observations
conducted at Arecibo are indicated by the shaded histogram.\label{FIG6}}
\end{figure}

The total integrated (corrected) and peak HI line
flux distributions are shown in Figure 7 and 8 respectively.  
Comparing these figures to the corresponding distributions for the HIPASS/HICAT
blind HI survey shown in Figure 9 of 
Meyer \etal ~(2004) and Figure 4 of Zwaan \etal ~(2004), it is readily 
apparent that the optically--targeted archive presented here contains
a substantial number of galaxies with both integrated fluxes and peak
fluxes well below those included in the HIPASS catalog. 
The HICAT sample contains very few objects with $S_{HI} <$ 2.5 Jy~\kms,
whereas almost one third of the optically selected sample presented here
have fluxes $S_c^{abs}$ lower than that value. Likewise, HICAT strives to be
peak flux limited and contains almost no sources with peak HI fluxes
below 30 mJy, whereas the majority of galaxies in the optically-selected
sample have peak fluxes below that value.

\begin{figure}
\figurenum{7}
\epsscale{1.0}
\plotone{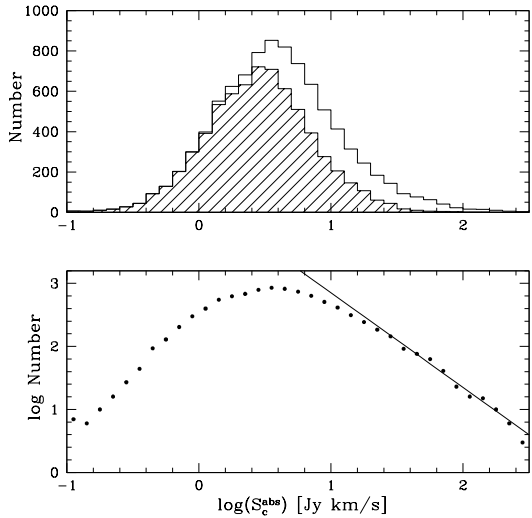}

\figcaption{Distribution of corrected HI fluxes $S_c^{abs}$ for all galaxies in 
the archive, in bins of width 0.1 dex. Observations
conducted at Arecibo are indicated by the shaded histogram.
The solid line in the {\it bottom} 
plot traces the case where log $N_{galaxies} \propto -1.5$ log $S_c^{abs}$, 
which is the dependence expected of an HI flux limited sample. .\label{FIG7}}
\end{figure}

\begin{figure}
\figurenum{8}
\epsscale{1.0}
\plotone{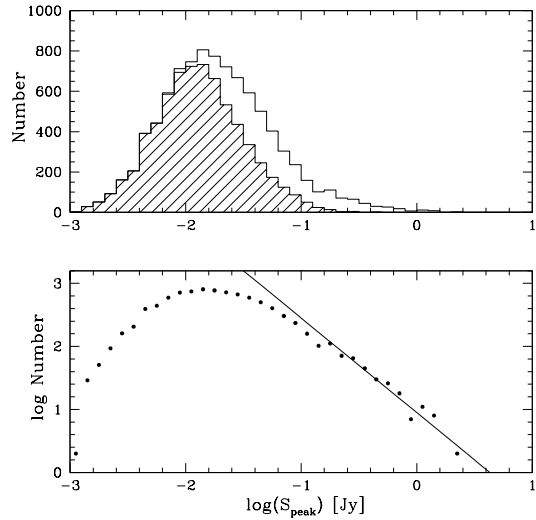}

\figcaption{Distribution of peak HI fluxes $S_{peak}$ for all galaxies
in the archive, in bins of width 0.1 dex. Observations
conducted at Arecibo are indicated by the shaded histogram.
The solid line in the {\it bottom} plot traces the
case of log $N_{galaxies} \propto -1.5$ log $S_{peak}$, 
which is the dependence expected of a peak flux limited sample.
The heterogenous nature of the observations that contribute to
this sample is very evident by the lack of a cutoff minimum $S_{peak}$.\label{FIG8}}
\end{figure}

For an HI flux limited sample, the logarithm of the number of galaxies per unit log $S_c^{abs}$ 
should be proportional to $-1.5~log~S_c^{abs}$; this relationship should also hold for the 
peak flux in a peak flux limited sample. The bottom panels of Figures 7 and 8 show the 
log-log distributions of corrected flux $S_c^{abs}$ and peak HI line flux
$S_{peak}$, with the $N \propto -1.5~
log ~S$ line superposed  for reference. Clearly the optically-selected sample is
neither peak HI flux nor total HI flux limited as the HI-blind surveys
strive to be. However, as shown in Springob, Haynes, \& Giovanelli (2005), with the addition of
supplemental data from the literature, it is possible to extract an optical diameter 
limited and HI flux limited subsample, restricted to the portion of
the sky lying within the Arecibo declination limits.

Figure 9 shows the distribution of instrumentally--corrected (but not
turbulence-- or inclination--corrected) widths $W_c$ 
for all galaxies in the archive. Objects with confused, asymmetric or
otherwise oddly behaved spectral profiles (coded ``C'', ``P'' or ``F'' in
Table 3) are included in the full distribution, but profiles which 
are classified as ``G'' or ``S'' are included in the shaded area. In
comparison to the HIPASS/HICAT dataset, the galaxies included in this
optically-selected sample are preferentially faster rotators,
unsuprising given the Tully--Fisher relation and the much larger volume
sampled here. This dataset is being used to derive the
observationally-based rotational velocity function for HI disk systems
(Springob, Haynes \& Giovanelli, in preparation).

\begin{figure}
\figurenum{9}
\epsscale{1.0}
\plotone{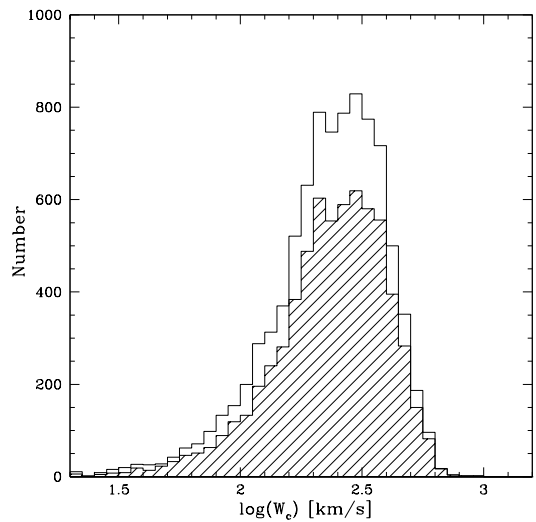}
\figcaption{Distribution of corrected HI widths $W_c$ for all galaxies in 
the archive except those identified as confused, in bins of width 0.05 dex. Observations 
conducted at Arecibo are indicated by the shaded histogram.
\label{FIG9}}
\end{figure}

There is one interesting issue for which much of this dataset is
unfortunately largely unsuited: the study of global asymmetry in HI profiles
(Richter \& Sancisi 1994; Haynes \etal ~1998). As discussed in Section
3.1, the shapes of global HI profiles observed with single dish telescopes
are affected by both beam attenuation and pointing offsets and errors.
Prior to its recent upgrade, the Arecibo telescope L-band feeds were
subject to pointing errors of typically 20--30\arcsec ~but sometimes
larger, especially when the observations were conducted at large
zenith angle. We therefore caution against using the spectra contained
in the AGC digital archive for studies of asymmetry, with the
exception of the very carefully collected Green Bank 42~m telescope
dataset obtained for that very purpose as discussed in Haynes \etal ~(1998).

\section {Summary}

We present here a compilation of HI spectral parameters for more than 9000 
optically-targeted galaxies in the local universe which have been
observed by us and our collaborators over the last 20$+$ years. All of
the HI spectra have been reprocessed in a homogenous manner and a
single set of algorithms has been used to extract spectral
parameters which can then be corrected to physical parameters taking
into account the instrument-- and processing--related 
complications such as spectral resolution, source extent, pointing
offsets and signal--to--noise. We use both empirical evidence and the
results of simulated observations to produce recipes for deriving
these physical parameters from the observed ones based on all of the
information available from the digital spectra themselves.

The recent blind HI surveys such as HIPASS offer the advantage of 
completeness and homogenous data processing and parameter extraction.
By reprocessing all of our existing HI line spectra using software
specifically designed to exploit the details of the observing setups
and the characteristics of the spectra themselves, we 
have tried to minimize the effects of instrument-- and processing--related
differences among the datasets. At the same time, HI
completeness issues remain in optically targeted samples, despite
their greater depth and sensitivity. It is well known that optical
galaxy catalogs often preferentially miss the low optical
surface brightness but HI-rich late type galaxies that are
common in the local universe. In contrast, blind HI surveys do not suffer
from this strong bias (Rosenberg \& Schneider 2002; Koribalski
\etal ~2004). A more thorough discussion of the completeness of the
AGC HI compilation in the context of the derivation of the HI
mass function is presented in Springob, Haynes \& Giovanelli (2005).
Despite the lack of flux completeness, 
the redshift, emission line flux, and observed rotational widths available
from the global HI line profiles collected here provide important 
measures of fundamental galaxy properties useful to diverse studies 
of galaxies in the nearby universe, especially when combined with 
complementary multiwavelength data.

In addition to the tabulations contained herein, 
the actual smoothed and baseline subtracted spectra are 
available in digital form by contacting the authors. Plans are
underway to make the entire dataset availble through the US National
Virtual Observatory in conjunction with a wide area blind HI survey
to be undertaken at Arecibo beginning in 2005. This {\it Arecibo
Legacy Fast ALFA} (ALFALFA) survey will make use of the new 7--feed
dual polarization, L-band array (ALFA) to survey over 7000 deg$^{2}$ of the sky
visible to Arecibo, detecting tens of thousands of HI sources, 
from local, very low HI mass dwarfs to gas-rich massive 
galaxies seen to $z \sim0.06$. A careful study of the detection
statistics of the present optically selected sample in such a 
sensitive HI blind survey will provide insight into the differential
characteristics of galaxies according to optical surface brightness and HI content.  Only the mining of datasets derived
from wide area surveys conducted at multiple wavelengths will, in
combination, provide the true census of all galaxies that occupy the
local universe. 

\vskip 0.3in

We wish to thank our numerous collaborators, colleagues, students,
telescope staff and friends who have participated over many years and
in many places and contexts in the acquisition
of the original data used to produce this archive.
We especially thank Barbara Catinella for conducting the fall 2001 Arecibo observations, 
and Joe McMullin, Bob Garwood, Karen O'Neil, and Jeremy Darling for their work in 
adapting AIPS++ for use with Arecibo spectral line data.  This work has been partially 
supported by AST-9900695, AST-0307396, AST-0307661, the N.R.A.O/GBT 03B-007
Graduate Student Support Grant
and the NASA New York State Space Grant.

\vskip 0.3in

{\bf Tables 3, 4, and 5} (in ASCII format) are exclusively available online at {\it http://egg.astro.cornell.edu/hiarchive/hiarchive.html}.

\vfill
\end{document}